\documentclass{emulateapj}

\newcommand{\vrot}{$v_{\rm rot}\sin{i}$}
\usepackage{multirow}
\usepackage{amssymb}
\newcommand{\ptf}{PTF1\,J0823}
\newcommand{\kms}{\ensuremath{{\rm km}\,{\rm s}^{-1}}}
\newcommand{\degree}{$^{\circ}$ }
\newcommand{\msol}{M$_\odot$}
\newcommand{\teff}{T$_{\rm eff}$}
\newcommand{\logg}{$\log{g}$}
\newcommand{\porb}{P$_{\rm orb}$}

\begin{document}


\title{PTF1\,J082340.04+081936.5: A hot subdwarf B star with a low mass white dwarf companion in an 87 minute orbit}


\author{Thomas Kupfer\altaffilmark{1,2}, Jan van Roestel\altaffilmark{3}, Jared Brooks\altaffilmark{4}, Stephan Geier\altaffilmark{5}, Tom R. Marsh\altaffilmark{6}, Paul J. Groot\altaffilmark{3}, Steven Bloemen\altaffilmark{3}, Thomas A. Prince\altaffilmark{2}, Eric Bellm\altaffilmark{2}, Ulrich Heber\altaffilmark{7}, Lars Bildsten\altaffilmark{4,8}, Adam A. Miller\altaffilmark{9,10}, Martin J. Dyer\altaffilmark{11}, Vik S. Dhillon\altaffilmark{11,12},  Matthew Green\altaffilmark{6}, Puji Irawati\altaffilmark{13}, Russ Laher\altaffilmark{14}, Stuart P. Littlefair\altaffilmark{11}, David L. Shupe\altaffilmark{15}, Charles C. Steidel\altaffilmark{2}, Somsawat Rattansoon\altaffilmark{11,13} and Max Pettini\altaffilmark{16}}






\altaffiltext{1}{tkupfer@caltech.edu}
\altaffiltext{2}{Division of Physics, Mathematics and Astronomy, California Institute of Technology, Pasadena, CA 91125, USA}
\altaffiltext{3}{Department of Astrophysics/IMAPP, Radboud University Nijmegen, PO Box 9010, NL-6500 GL Nijmegen, The Netherlands}
\altaffiltext{4}{Department of Physics, University of California, Santa Barbara, CA 93106, USA}
\altaffiltext{5}{Institute for Astronomy and Astrophysics, Kepler Center for Astro and Particle Physics, Eberhard Karls University, Sand 1,
72076 T\"ubingen, Germany}
\altaffiltext{6}{Department of Physics, University of Warwick, Coventry CV4 7AL, UK}
\altaffiltext{7}{Dr. Remeis-Sternwarte \& ECAP, Astronomical Institute, University of Erlangen-Nuremberg, Germany}
\altaffiltext{8}{Kavli Institute for Theoretical Physics, Santa Barbara, CA 93106, USA}
\altaffiltext{9}{Center for Interdisciplinary Exploration and Research in Astrophysics (CIERA) and Department of Physics and Astronomy, Northwestern University, 2145 Sheridan Road, Evanston, IL 60208, USA}
\altaffiltext{10}{The Adler Planetarium, 1300 S. Lakeshore Drive, Chicago, IL 60605, USA}
\altaffiltext{11}{Department of Physics \& Astronomy, University of Sheffield, Sheffield, S3 7RH, UK}
\altaffiltext{12}{Instituto de Astrofísica de Canarias (IAC), E-38200 La Laguna, Tenerife, Spain}
\altaffiltext{13}{National Astronomical Research Institute of Thailand, 191 Siriphanich Building, Huay Kaew Road, Chiang Mai 50200, Thailand}
\altaffiltext{14}{Spitzer Science Center, California Institute of Technology, Pasadena, CA 91125, USA}
\altaffiltext{15}{Infrared Processing and Analysis Center, California Institute of Technology, Pasadena, CA 91125, USA}
\altaffiltext{16}{Institute of Astronomy, Madingley Road, Cambridge CB3 0HA, UK}

\begin{abstract}


We present the discovery of the hot subdwarf B star (sdB) binary PTF1\,J082340.04+081936.5. The system has an orbital period \porb=87.49668(1)\,min (0.060761584(10) days), making it the second-most compact sdB binary known. The lightcurve shows ellipsoidal variations. Under the assumption that the sdB primary is synchronized with the orbit, we find a mass $M_{\rm sdB}=0.45^{+0.09}_{-0.07}$\,\msol, a companion white dwarf mass $M_{\rm WD}=0.46^{+0.12}_{-0.09}$\,\msol\,and a mass ratio $q = \frac{M_{\rm WD}}{M_{\rm sdB}}=1.03^{+0.10}_{-0.08}$. 

The future evolution was calculated using the \texttt{MESA} stellar evolution code. Adopting a canonical sdB mass of $M_{\rm sdB}=0.47$\,\msol\, we find that the sdB still burns helium at the time it will fill its Roche lobe if the orbital period was less than 106\,min at the exit from the last common envelope phase. For longer common envelope exit periods the sdB will have stopped burning helium and turned into a C/O white dwarf at the time of contact. Comparing the spectroscopically derived \logg\,and $T_{\rm eff}$ with our \texttt{MESA} models, we find that an sdB model with a hydrogen envelope mass of $5\times10^{-4}\,M_\odot$ matches the measurements at a post-common envelope age of 94\,Myr, corresponding to a post-common envelope orbital period of 109\,min which is close to the limit to start accretion while the sdB is still burning helium.



\end{abstract}




\section{Introduction}\label{sec:intro}
Hot subdwarf B stars (sdBs) are core helium burning stars with masses around 0.5 \msol\, and thin hydrogen envelopes (\citealt{heb86,heb09,heb16}). A large fraction of sdBs are members of short-period binaries with periods below $\approx10$\,days \citep{max01,nap04a}. Orbital shrinkage through a common envelope (CE) phase is the only likely formation channel for such short period sdB binaries \citep{han02,han03}. 



\begin{table*}[t]
 \centering
 \caption{Summary of the observations of \ptf}
  \begin{tabular}{cclccl}
  \hline\hline
 Date &    UT  &  Tele./Inst. & N$_{\rm exp}$ & Exp. time (s) & Coverage (\AA)/Filter \\
  \hline
    {\bf Photometry} &        &        \\
  2009-11-16 - 2015-01-15  &    & Palomar 48-inch  &  144  & 60 &  R$_{\rm mould}$ \\
  2016-01-31  & 14:10 - 16:50    & TNT/ULTRASPEC & 2404 & 3.94 & $g^\prime$ \\
     \noalign{\smallskip}
 {\bf Spectroscopy} &     &           \\
  2015-10-25  &  11:30 -  12:00 &  200-inch/DBSP & 7 & 240    &  3800 - 10\,000 \\    \vspace{-0.08cm} 
  \multirow{2}{*}{2016-01-30}  &  07:04 -  07:52  & \multirow{2}{*}{200-inch/DBSP}  & \multirow{2}{*}{16} & \multirow{2}{*}{240}    & \multirow{2}{*}{3800 - 10\,000}  \\ 
                & 09:42 - 10:03   &     &   &    & \\
  2016-02-03  &  06:59 -  08:26  & 200-inch/DBSP  & 20 & 240    & 3800 - 10\,000  \\
  2016-03-01  &    05:46 - 06:07  & Keck/HIRES    &  6 & 180       &  3700 - 5000 \\ 
  2016-04-13  &   20:18 - 21:21    & WHT/ISIS       &  12 & 290       &  3700 - 7400  \\ 
   \hline
\end{tabular}
\label{observ}
\end{table*}


Evolutionary studies have shown that the orbital period of a hot subdwarf with a white dwarf companion has to be \porb$\lesssim$\,120\,min on exit from the last CE to still have an sdB that is core or shell helium burning when it fills its Roche lobe assuming that the further orbital period evolution is set by the emission of gravitational waves only. In the subsequent evolution, if helium burning is still ongoing, the sdB fills its Roche lobe first and starts to transfer He-rich material onto the white dwarf (e.g. \citealt{tut89,tut90,ibe91,yun08}). If the system has a mass ratio $q = M_{\rm sdB}/M_{\rm WD}\,\lesssim2$, stable mass-transfer is possible \citep{sav86,wan09}. Mass transfer starts at orbital periods ranging from 16 to 50 min. Subsequently the semi-detached system evolves to shorter periods with typical mass transfer rates of $\dot{M}\approx10^{-8}\,$\msol yr$^{-1}$ (e.g. \citealt{sav86,yun08,pie14,bro15}). 

CD$-$30$^\circ$11223 has the shortest known orbital period of all sdB binaries (\porb\,= 70.52986\,min) and is the only known system where the sdB is still expected to be burning helium when it will fill its Roche lobe \citep{ven12, gei13}. After accreting 0.1\,\msol, He-burning is predicted to be ignited unstably in the accreted helium layer. This in turn triggers the ignition of carbon in the core which might disrupt the WD even when the mass is significantly below the Chandrasekhar mass (e.g. \citealt{liv90,liv95,fin10,woo11,gei13,she14}). If the WD is not disrupted, the unstable burning of the He-shell will detonate the shell and may be observed as a faint and fast .Ia supernova \citep{bil07}. This increases the orbital separation, but gravitational wave radiation drives the donor back into contact, resuming mass transfer and triggering several subsequent weaker flashes \citep{bro15}.

Inspired by the discovery of CD$-$30$^\circ$11223, we have conducted a search for (ultra-)compact post-common envelope systems using the Palomar Transient Factory (PTF; \citealt{law09,rau09}) large area synoptic survey based on a color selected sample from the Sloan Digital Sky Survey (SDSS). The PTF uses the Palomar 48$^{\prime\prime}$ Samuel Oschin Schmidt telescope to image up to $\approx2000$\,deg$^2$ of the sky per night to a depth of R$_{\rm mould} \approx20.6$\,mag or $g' \approx21.3$\,mag. Here we report the discovery of the ultracompact sdB+WD system PTF1\,J082340.04+081936.5 (hereafter PTF1\,J0823). PTF1\,J0823 has the second shortest orbital period amongst the known binaries with a hot subdwarf component. 
  

\section{Observations and Data reduction}\label{obser}

\subsection{Photometry}

As part of the Palomar Transient Factory, the Palomar 48-inch (P48) telescope images the sky every night. The reduction pipeline for PTF applies standard de-biasing, flat-fielding, and astrometric calibration to raw images \citep{lah14}. Relative photometry correction is applied and absolute photometric calibration to the few percent level is performed using a fit to SDSS fields observed in the same night \citep{ofe12}. The lightcurve of \ptf\, has 144 epochs with good photometry in the R$_{\rm mould}$ band with a typical uncertainty of 0.01 mag. The cadence is highly irregular, ranging from 10 minutes to years with an average separation of about $10$\,days.

High cadence observations were conducted using the 2.4-m Thai National Telescope (TNT) with the high-speed photometer ULTRASPEC \citep{dhi14}. ULTRASPEC employs a 1024x1024 pixel frame-transfer, electron-multiplying CCD (EMCCD) in conjunction with re-imaging optics to image a field of 7.7$^\prime$x7.7$^\prime$ at (windowed) frame rates of up to $\sim$\,200 Hz. 
Observations were obtained with the $g^\prime$ filter on January 31 2016 over 1h50min with an exposure time of 3.9\,sec and a dead time of 15\,ms leading to a total of 2404 epochs. Data reduction was carried out with the ULTRACAM pipeline \citep{dhi07}. All frames were bias-subtracted and flat-fielded.

\subsection{Spectroscopy}
Optical spectra of \ptf\, were obtained with the Palomar 200-inch telescope and the Double-Beam Spectrograph (DBSP; \citealt{oke82}) over 3 nights using a low resolution mode ($R\sim1500$).  We took 7 consecutive exposures on October 25 2015, 16 exposures on January 30 2016 and 20 exposures on February 2 2106 all with an exposure time of $240$\,sec.  Each night an average bias frame was made out of 10 individual bias frames and a normalized flat-field frame was constructed out of 10 individual lamp flat-fields. For the blue arm, FeAr and for the red arm, HeNeAr arc exposures were taken in the beginning and end of the night. Each exposure was wavelength calibrated by interpolating between the beginning and end of the night calibration exposures. Both arms of the spectrograph were reduced using a custom \texttt{PyRAF}-based pipeline \footnote{https://github.com/ebellm/pyraf-dbsp}\citep{bel16}. The pipeline performs standard image processing and spectral reduction procedures, including bias subtraction, flat-field correction, wavelength calibration, optimal spectral extraction, and flux calibration. 

On April 13 2016, we obtained 12 consecutive spectra using the William Herschel Telescope (WHT) and the ISIS spectrograph \citep{car93} using a medium resolution mode (R600B grating, $R\approx3000$). One hundred bias frames were obtained to construct an average bias frame and 100 individual tungsten lamp flat-fields were obtained to construct a normalized flat-field. \mbox{CuNeAr} arc exposures were taken before and after the observing sequence as well as after 6 spectra to correct for instrumental flexure. Each exposure was wavelength calibrated by interpolating between the two closest calibration exposures. All spectra were de-biased and flat fielded using \texttt{IRAF} routines. One dimensional spectra were extracted using optimal extraction and were subsequently wavelength and flux calibrated.

Additionally, \ptf\,was also observed on March 1 2016 using Keck with the HIRES spectrograph ($R\approx36\,000$). The full data set consists of 6 spectra which were taken consecutively. ThAr arc exposures were taken at the beginning of the night. The spectra were reduced using the \texttt{MAKEE}\footnote{http://www.astro.caltech.edu/$\sim$tb/ipac$\_$staff/tab/makee/} pipeline following the standard procedure: bias subtraction, flat fielding, sky subtraction, order extraction, and wavelength calibration.

\begin{figure}
\begin{center}
\includegraphics[width=0.49\textwidth]{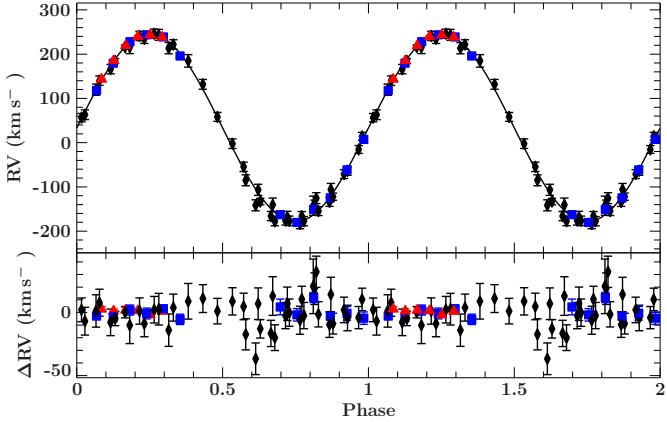}
\end{center}
\caption{Radial velocity plotted against orbital phase. The RV data were phase folded with the orbital period and are plotted twice for better visualization. The residuals are plotted below. The RVs were measured from spectra obtained with P200/DBSP (black diamonds), WHT/ISIS (blue squares) and Keck/HIRES (red triangles).}
\label{fig:rv_curve}
\end{figure}

\begin{figure*}
\begin{center}
\includegraphics[width=0.98\textwidth]{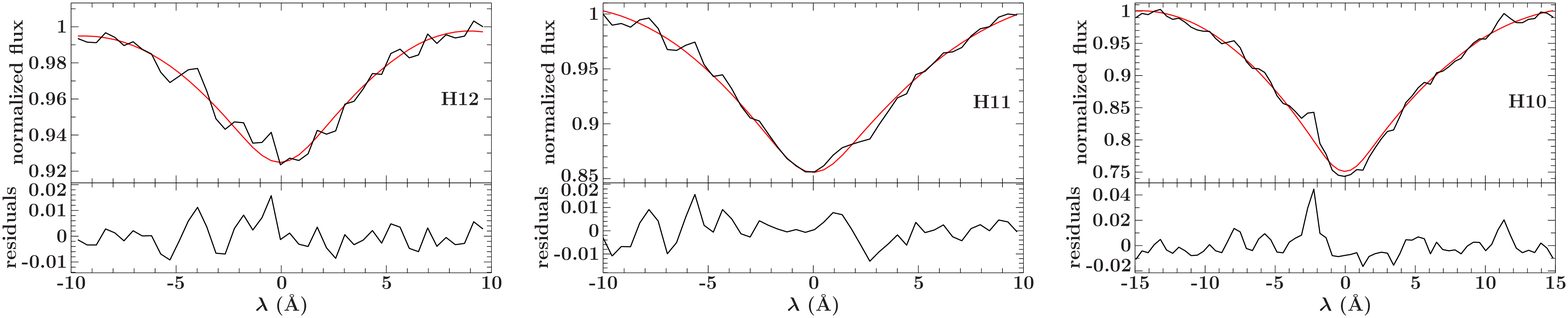}
\includegraphics[width=0.98\textwidth]{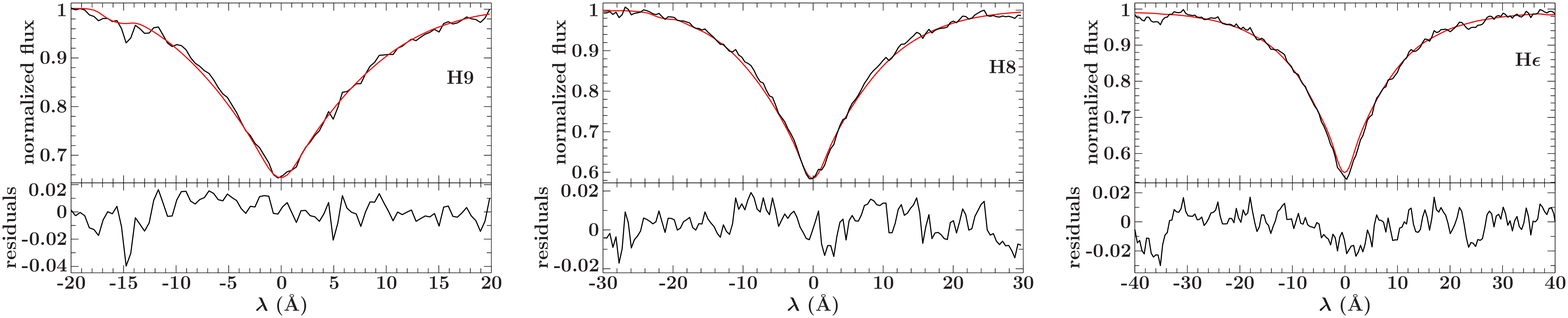}
\includegraphics[width=0.98\textwidth]{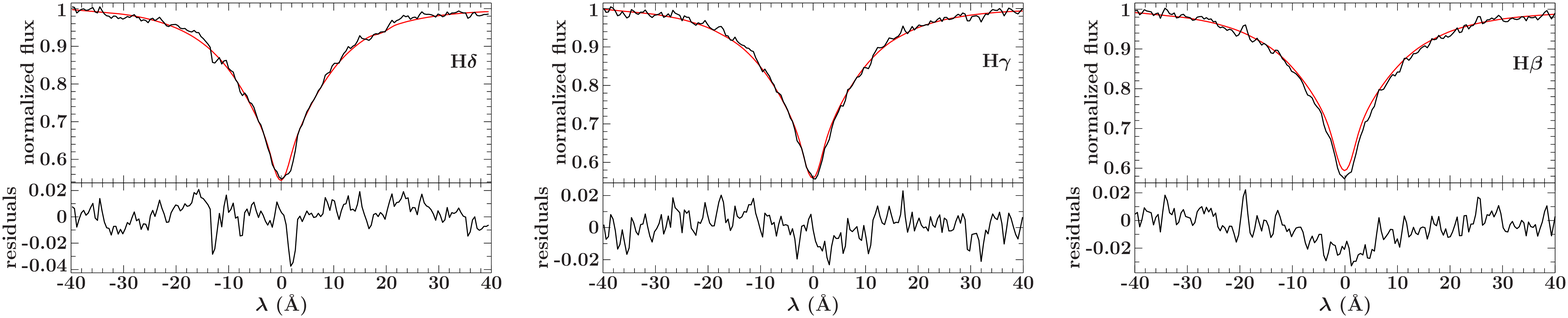}
\end{center}
\caption{Fit of synthetic LTE models to the hydrogen Balmer lines of a coadded ISIS spectrum. The normalized fluxes of the single lines are shifted for better visualisation.}
\label{fig:wht_model}
\end{figure*} 

Table\,\ref{observ} gives an overview of all observations and the instrumental set-ups.

\section{Orbital and atmospheric parameters}\label{orb_atm_pars}
The dominant variation in lightcurve is due to the ellipsoidal deformation of the sdB primary. This is caused by the tidal influence of the compact companion. The lightcurve also shows Doppler boosting, caused by the orbital motion of the sdB \citep{sha87,blo11,gei13}. The ephemeris has been derived from the PTF lightucrve using the \texttt{Gatspy} module for time series analysis which uses the Lomb-Scargle periodogram\footnote{http://dx.doi.org/10.5281/zenodo.14833}\citep{van15}. Because of the timebase of more than five years, the derived orbital period of \porb=87.49668(1)\,min (0.060761584(10) days) is accurate to 1\,ms. The error was estimated by bootstrapping the data. 

To obtain radial velocities, we followed the procedure as described in detail in \citet{gei11a}. To fit the continuum, line and line core of the individual lines we fitted Gaussians, Lorentzians and polynomials to the hydrogen and helium lines using the \texttt{FITSB2} routine \citep{nap04a}. The wavelength shifts compared to the rest wavelengths of all suitable spectral lines were fitted simultaneously using a $\chi^2$-minimization. 
We found consistent velocities between the 6 HIRES spectra and the ISIS spectra taken at the same orbital phase, as well as consistent velocity amplitudes between the ISIS and DBSP spectra. However, we found a significant offset of the systematic velocities between the DBSP and the ISIS spectra. In the night of January 30 2016 the systemic velocity of the DBSP spectra was shifted by 50\,\kms. The DBSP calibration spectra were taken at the beginning and end of the night. For ISIS the calibration lamps where taken at the position of the target before, after and during the sequence and the velocities from the HIRES spectra are consistent with the ISIS spectra. Therefore, we conclude that the offset in the DBSP spectra is due to instrumental flexure because the calibration lamps were not taken at the position of the object. We corrected the velocities measured in the DBSP spectra to fit the systemic velocity obtained by the ISIS spectra. All velocities were folded on the ephemeris which was derived from the photometric data. Assuming circular orbits, a sine curve was fitted to the folded radial velocity (RV) data points. We find a semi-amplitude K$=211.7\pm1.8$\,\kms\, and a systemic velocity of $\gamma=33.3\pm1.4\,$\kms (Fig.\,\ref{fig:rv_curve}).

\begin{figure}
\begin{center}
\includegraphics[width=0.49\textwidth]{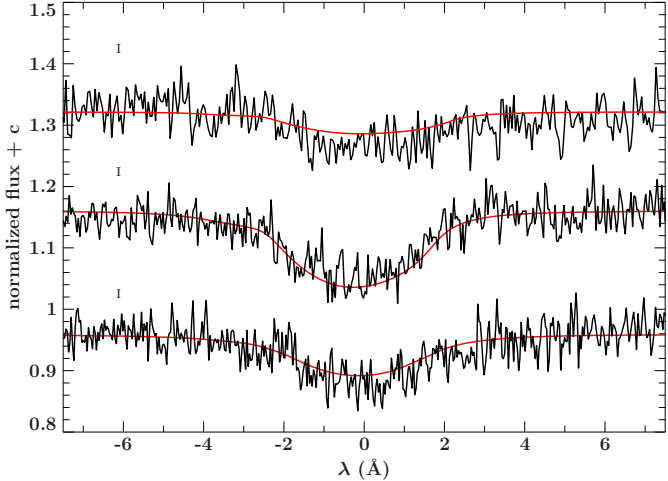}
\end{center}
\caption{Best fits of \vrot\,to the helium lines seen in the HIRES spectra. The atmospheric parameters were fixed to the values derived from the WHT spectra.}
\label{fig:rotation}
\end{figure}

The atmospheric parameters of effective temperature, \teff, surface gravity, \logg, helium abundance, $\log{y}$, and projected rotational velocity, \vrot, were determined by fitting the rest-wavelength corrected average DBSP, ISIS and HIRES spectra with metal-line-blanketed LTE model spectra \citep{heb00}. The most sensitive lines for \logg~and \teff~are the Balmer lines close to the Balmer jump. We used the hydrogen lines H12 ($3750$\AA) up to H$\beta$ in the WHT/ISIS spectrum to measure \teff~and \logg~with \vrot~as a free parameter and found \teff=27\,100$\pm$500\,K, \logg=5.50$\pm$0.05 and \vrot=122$\pm$21\,\kms (Fig.\,\ref{fig:wht_model}). The HIRES spectra are not well suited to measure \teff~and \logg~because the broad hydrogen absorption lines span several orders and merging of the echelle spectra could introduce systematic errors. However, the high-resolution HIRES spectra are well suited to measure the projected rotational velocity \vrot~of the sdB in lines which are not affected by order merging. The three helium lines ($4026, 4471, 4921$\,\AA) which are covered by the HIRES spectrum and not affected by order merging are less sensitive to \teff~and \logg, and most sensitive to rotational broadening \vrot~ and the helium abundance $\log{y}$. Therefore, they were used to measure $\log{y}$~and \vrot, keeping \teff~and \logg~fixed to the values measured from the ISIS spectra. We found \vrot=132$\pm$5\,\kms and $\log{y}$=$-$2.47$\pm$0.03 (Fig.\,\ref{fig:rotation}). As a consistency check \teff~and \logg~were derived from the DBSP spectrum keeping $\log{y}$~and \vrot~fixed to the values measured from HIRES. Although, the DBSP spectrum only covers hydrogen lines down to H9 ($3835$\,\AA) we find good agreement within the errors with the parameters derived from the ISIS spectra. However, because of the larger coverage of Balmer lines, the further analysis will be done using \teff~ and \logg~ from the ISIS spectra.

Table\,\ref{tab:atmo} shows the atmospheric parameters and Table\,\ref{tab:system} summarizes the orbital parameters.





\section{Lightcurve analysis}

To model the lightcurve, we used the \texttt{LCURVE} code \citep{cop10}. \texttt{LCURVE} uses many points in a grid to model the surfaces of the stars with a shape according to the Roche geometry. We assume co-rotation and an eccentricity of zero. The flux that each point on the grid emits is calculated by assuming a blackbody of a certain temperature at the bandpass wavelength, corrected for limb darkening, gravity darkening, Doppler beaming and the reflection effect.
\begin{table}[t]
 \centering
 \caption{Summary of atmospheric parameter \ptf}
  \begin{tabular}{lllll}
  \hline\hline
  Telescope &    \teff\,(K)  &  \logg & $\log{y}$ & \vrot \\
                    & (K)     &      &     &   (\kms)  \\
  \hline
  DBSP  &     26\,700$\pm$600  &  5.49$\pm$0.06   &   $-$2.47$^a$   & 132$^a$ \\
  ISIS      &    27\,100$\pm$500   &  5.50$\pm$0.05    &  $-$2.47$^a$   & 122$\pm$21   \\
  HIRES        &   27\,100$^b$     &  5.50$^b$   &   $-$2.47$\pm$0.03  &  132$\pm$5  \\
   \hline
   adopted & 27\,100$\pm$500  & 5.50$\pm$0.05  &  $-$2.47$\pm$0.03  &  132$\pm$5  \\
   \hline
\end{tabular}
\label{tab:atmo}
\begin{flushleft}
$^a$ fixed to the values derived from HIRES\\
$^b$ fixed to the values derived from ISIS
\end{flushleft}
\end{table}

We use information from spectroscopy and the P48 lightcurve to fix or constrain some of the model parameters. First, we fix the orbital period to the value as determined in section \ref{orb_atm_pars}. Second, we fix the primary temperature (\teff), primary radial velocity amplitude ($K$), the surface gravity of the primary ($g$) and the rotational velocity (\vrot, see section \ref{orb_atm_pars}). As an additional constraint we use as a lower limit for the white dwarf radius the zero-temperature mass radius relation by Eggleton (quoted from \citealt{ver88}). We use the same method to account for limb darkening and gravity darkening as described in \citet{blo11}: the Claret limb darkening prescription \cite{cla04} and the gravity darkening prescription from \citet{zei24} with a passband specific gravity darkening coefficient. We investigated how the limb darkening coefficient affects the results by adding it as free parameter with $\mathrm{\beta}=0.1 - 1.0$. The co-variance between the limb darkening parameter and system parameters is negligible compared to the uncertainty on the parameters. Therefore, we kept the limb darkening coefficients fixed for the analysis. The limb darkening coefficients are $\mathrm{a_1}=0.677$,  $\mathrm{a_2}=-0.312$, $\mathrm{a_3}=0.212$, and $\mathrm{a_4}=-0.079$ for the limb darkening coefficient $\mathrm{\beta}=0.460$. These values are calculated for an sdB with a temperature \teff=27\,100\,K and \logg=5.50 using the models from \citet{blo11}. We did not use any limb darkening or gravity darkening in the white dwarf model, since these do not affect the light curve. This leaves as free parameters in the model the mass ratio $q$, the inclination $i$, secondary temperature $T_{\rm WD}$, the scaled radii of both components $r_{\rm sdB,WD}$, the velocity scale ($\mathrm[K+K_{\rm WD}]/\sin i$) and the beaming parameter $B$ ($F_\lambda = F_{0,\lambda} \lbrack 1 - B \frac{v_r}{c}\rbrack$, see \citealt{blo11}). Besides these system parameters we add a third order polynomial to correct for any residual airmass effects. 
 
To determine the uncertainties on the parameters we combine \texttt{LCURVE} with \texttt{emcee} \citep{for13}. \texttt{emcee} is an implementation of an MCMC sampler and uses a number of parallel chains to explore the solution space. We use 2048 chains and let them run until the chains stabilized to a solution, which took approximately 6000 generations. 

\begin{table*}[t]
\centering
\caption{Overview of the derived parameter for \ptf}
\begin{tabular}{lll}
\hline\hline
Right ascension &  RA [hrs]  & 08:23:40.04 \\
Declination  &  Dec $[^\circ]$  & +08:19:36.5 \\
Visual magnitude$^a$ & $m_{\rm V}$ & 14.681$\pm$0.051 \\
\hline
\multicolumn{3}{l}{\bf{Atmospheric parameter of the sdB}}  \\ 
Effective temperature & \teff\,[K] & 27100$\pm$500 \\
Surface gravity   & \logg  & 5.50$\pm$0.05  \\
Helium abundance& $\log{y}$  & --1.47$\pm$0.03 \\
Projected rotational velocity & \vrot\,[\kms] & 132$\pm$5 \\
\hline
\multicolumn{3}{l}{\bf{Orbital parameter}}   \\ 

&  $T_0$ [BJD UTC]  & 57418.6202(2)  \\
Orbital period & \porb\,[d]  & 87.49668(1) \\
RV semi-amplitude & $K$ [\kms] & 211.7$\pm$1.8 \\
System velocity & $\gamma$\,[\kms] & 33.3$\pm$1.4\\ 
Binary mass function & $f_{\rm m}$ [\msol] &  0.0597$\pm$0.0020 \\
\hline
\multicolumn{3}{l}{\bf{Derived parameter}} \\

Mass ratio  &  $q = \frac{M_{\rm WD}}{M_{\rm sdB}}$  & $1.03^{+0.10}_{-0.08}$  \\
sdB mass &  $M_{\rm sdB}$ [\msol] & $0.45^{+0.09}_{-0.07}$ \\ 
sdB radius & $R_{\rm sdB}$ [R$_{\odot}$] &  $0.20^{+0.03}_{-0.02}$  \\ 
WD mass &  $M_{\rm WD}$ [\msol] & $0.46^{+0.12}_{-0.09}$ \\
Orbital inclination & $i\,[^\circ$] & $52^{+8}_{-7}$   \\
Separation  & $a$ [R$_{\odot}$]   &  $0.63^{+0.05}_{-0.04}$ \\
Distance & $d$ [kpc] & $1.2^{+0.2}_{-0.2}$ \\
\hline
\multicolumn{3}{l}{\bf{Derived parameter assuming the canonical sdB mass}} \\

Mass ratio  &  $q = \frac{M_{\rm WD}}{M_{\rm sdB}}$  & $1.04^{+0.10}_{-0.08}$ \\
sdB mass & $M_{\rm sdB}$ [\msol] & 0.47 (fixed) \\ 
sdB radius & $R_{\rm sdB}$ [R$_{\odot}$] & $0.204^{+0.007}_{-0.006}$ \\ 
WD mass & $M_{\rm WD}$ [\msol] & $0.49^{+0.05}_{-0.04}$  \\
Orbital inclination & $i\,[^\circ$] & $51^{+4}_{-4}$  \\
Separation  &  $a$ [R$_{\odot}$]   & $0.641^{+0.010}_{-0.009}$ \\
Distance & $d$ [kpc] & $1.2^{+0.1}_{-0.1}$ \\
\hline
\end{tabular}
\begin{flushleft}
\centering
$^a$ taken from the APASS catalog \citep{hen16}
\label{tab:system}
\end{flushleft}
\end{table*}

\begin{figure}
\begin{center}
\includegraphics[width=0.49\textwidth]{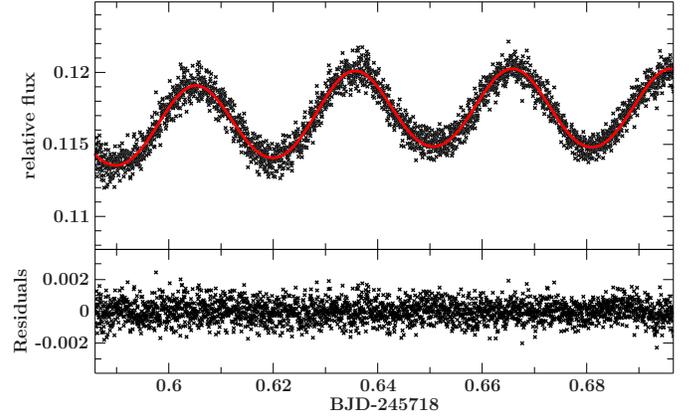}
\end{center}
\caption{lightcurve obtained with ULTRASPEC shown together with the \texttt{Lcurve} fit. The residuals are plotted below.}
\label{fig:lightcurve}
\end{figure}

With the surface gravity ($g$) and projected rotational velocity (\vrot), we have three equations at hand that constrain the system, with the sdB mass M$_{\rm sdB}$ the only free parameter. The binary mass function
 \begin{equation}
 \label{equ:mass-function}
 f_{\rm m} = \frac{M_{\rm WD}^3 \sin^3(i)}{(M_{\rm WD} +  M_{\rm sdB})^2} = \frac{P_{\rm orb} K^3}{2 \pi G}
 \end{equation}
can be combined with 
\begin{equation}
\sin i = \frac{(v_{\rm rot}\sin{i}) P_{\rm orb}}{2 \pi R_{\rm sdB}}    
\end{equation}
and
\begin{equation}
R_{\rm sdB}=\sqrt{\frac{M_{\rm sdB}G}{g}}
\end{equation}
with \porb\, being the orbital period, $K$ the velocity semi-amplitude, $M_{\rm WD}$ the mass of the companion and $R_{\rm sdB}$ the radius of the sdB. The approach is described in full detail in \citet{gei07}. A strict lower mass limit for the sdB can be derived, because the inclination cannot exceed $90^\circ$. We found a lower limit for the sdB mass M$_{\rm sdB}>0.25$\,\msol. The error is dominated by the surface gravity, which has to be estimated from the model atmosphere fit. 

\section{System parameters}
Because the system is not eclipsing we cannot obtain a unique solution for the component masses from the lightcurve analysis. In order to determine masses and radii of both the sdB and the WD companion, we combined the results from the lightcurve analysis with the assumption of tidal synchronization of the sdB primary to the orbit. The given errors are all $95\,\%$ confidence limits.

\begin{figure}
\begin{center}
\includegraphics[width=0.49\textwidth]{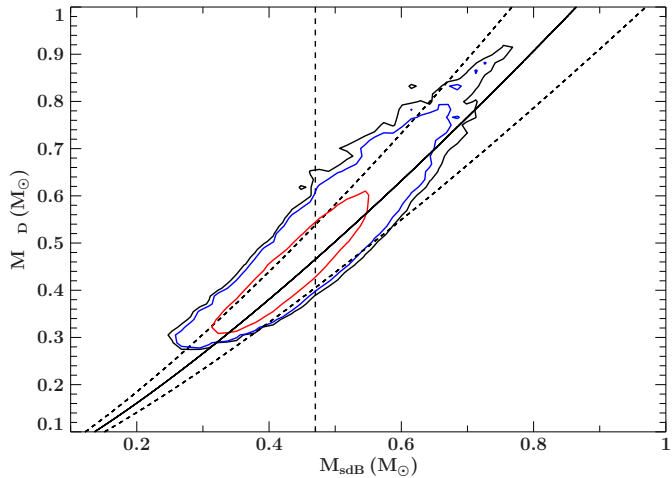}
\end{center}
\caption{White dwarf mass versus sdB mass. The curved lines correspond to synchronization with the corresponding error. The dashed vertical line marks the canonical sdB mass of $0.47$\,\msol. The contours show the results from the lightcurve fit with 1$\sigma$ (red), 2$\sigma$ (blue), 3$\sigma$ (black) confidence.}
\label{fig:companion_cont}
\end{figure}

\begin{figure}
\begin{center}
\includegraphics[width=0.49\textwidth]{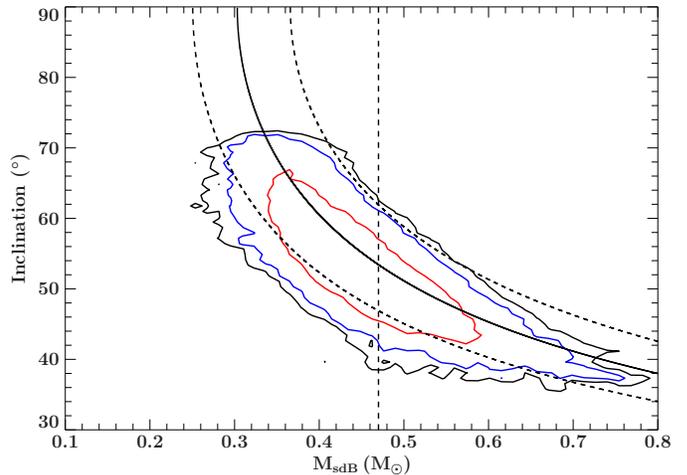}
\end{center}
\caption{Inclination versus sdB mass. The curved lines correspond to synchronization with the corresponding error. The dashed vertical line marks the canonical sdB mass of $0.47$\,\msol. The contours show the results from the lightcurve fit with 1$\sigma$ (red), 2$\sigma$ (blue), 3$\sigma$ (black) confidence.}
\label{fig:inclination_cont}
\end{figure}

We find that both components have nearly the same mass. A mass ratio $q = M_{\rm WD}/M_{\rm sdB}=1.03^{+0.10}_{-0.08}$, a mass for the sdB  $M_{\rm sdB}=0.45^{+0.09}_{-0.07}$\,\msol\,and a WD companion mass $M_{\rm WD}=0.46^{+0.12}_{-0.09}$\,\msol\, were derived (Fig.\,\ref{fig:companion_cont}). The inclination is found to be $i = 52^{+8}_{-7}$\,\degree\,(Fig.\,\ref{fig:inclination_cont}). The beaming factor is $B=1.3^{+0.4}_{-0.4}$, which is consistent with the theoretical value 1.74 (for an sdB with \teff$=27\,100$, \logg$=5.50$). Because the system is not eclipsing, the radius and temperature of the white dwarf are poorly constrained.

The distance to \ptf\, was calculated using the visual V magnitude ($m_{\rm V}$), the sdB mass, \teff\, and \logg\, as described in \citet{ram01}. We find a distance to \ptf\, of $d=1.2^{+0.2}_{-0.2}$kpc.

An overview of the derived system parameter is given in Table\,\ref{tab:system}.

\section{Discussion} 
 



\subsection{Evolutionary history}

In the standard scenario, the so-called 2nd common envelope channel, the system starts as a low mass binary with $\approx$1\,\msol\,components. The initially more massive star first evolves to become a WD. Subsequently, the sdB progenitor fills its Roche lobe at the tip of the red-giant branch (RGB), forming an sdB with a canonical mass of $M_{\rm WD}=0.47$\,\msol, set by the helium core flash, with a WD companion \citep{han02,han03}. \citet{han02} showed that the binding energy of the envelope is very small at the tip of the RGB for a $1$\,\msol\, star and therefore the orbital shrinkage in the CE phase is not significant. They predict that sdB+WD binaries are formed with orbital periods longer than found in \ptf.

In a different picture an ultracompact sdB+WD binary can also be formed from a more massive main-sequence binary where the sdB progenitor is $>2$\msol. This sdB progenitor ignites helium non-degenerately in the core and fills its Roche lobe during the Hertzsprung gap or at the base of the RGB, resulting in an sdB with either a lower or higher mass compared to the tip of the RGB \citep{nel10a,gei13}. In such systems, the envelope is more tightly bound and the orbital shrinkage required to eject the CE becomes higher \citep{nel10a,gei13}. \citet{gei13} showed that CD$-$30$^\circ$11223 evolved most likely from a $2$\,\msol\,progenitor for the sdB with a $3$ - $4$\,\msol\,companion for the WD progenitor. The WD companion in \ptf\, is, with an upper mass limit of $M_{\rm WD}=0.58$\,\msol, less massive than in CD$-$30$^\circ$11223 . 

Perhaps the most similar system is KPD\,0422+5421 \citep{koe98}. However, its orbital period is \porb=$129.6$\,min and therefore about $42$\,min longer than \ptf. Due to the longer period in KPD\,0422+5421, this system is easier to explain by the 2nd common envelope channel from \citet{han02}. 


\subsection{Future evolution}
\begin{figure}
\begin{center}
\includegraphics[width=0.49\textwidth]{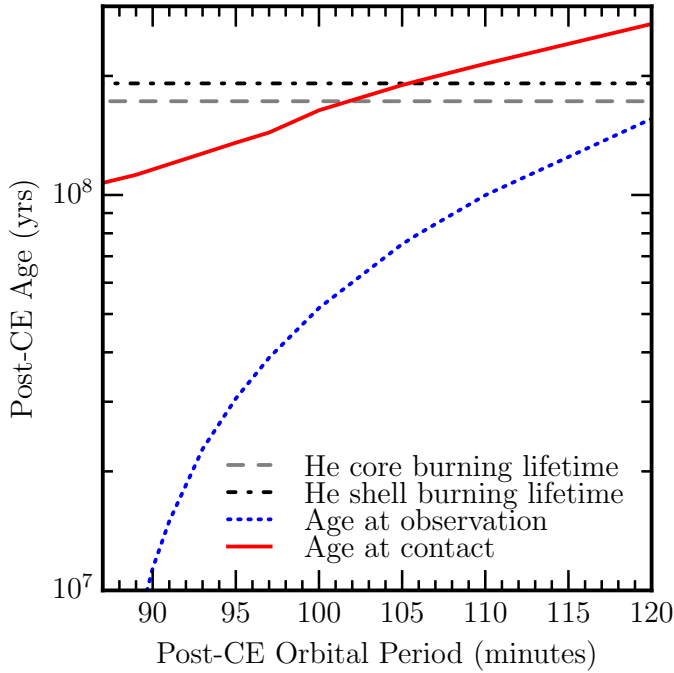}
\end{center}
\caption{The model post-CE orbital period is shown on the x-axis, with the post-CE age on the y-axis.
The dashed grey and dashed-dotted black lines show the core and shell He burning lifetimes of the sdB, respectively.
The dotted blue curve shows the ages of the models when the orbital period matches the period that we observed for this system.
The red curve shows the ages at which the stars make contact. 
Systems with initial orbital periods longer than 106 minutes will make contact after the sdB finishes He burning.}
\label{fig:evolution}
\end{figure}

To understand the future evolution of the system, we used the code \texttt{MESA} \citep{pax11,pax13,pax15}. 
For the model we assumed an sdB with a canonical mass \mbox{$M_{\rm sdB}=0.47$\,M$_\odot$} with a white dwarf companion of $M_{\rm WD}=0.49$\,\msol. 
Using release version 8118, we construct binary simulations that model the full stellar structure equations for the sdB and treat the WD as a point mass.
We ran a set of simulations with periods, when the system exits the CE (post-CE orbital period), ranging from 87 to 120 minutes.
The evolution of the system is governed by the loss of angular momentum due to radiation of gravitational waves.
We record the post-CE age at which the orbital periods match the observed period of 87 minutes, shown by the dotted blue curve in Fig. \ref{fig:evolution}, and the age at which the stars make contact, shown by the dashed-dotted red curve.
Recent modeling of asteroseismic observations of sdB stars \citep{con15} points to convective cores much larger than found with the Schwarzschild condition. At present, there is no clear consensus on the physics needed to achieve these larger cores, which prolongs the lifetime of the He burning phase. To accommodate such an outcome, we performed runs with element diffusion active \citep{mic07, sch15}, doubling the convective core mass (from $0.109 M_\odot$ to $0.218 M_\odot$) and the core-burning lifetime (from 80 Myr to 152 Myr). The data from these runs are shown in Figures \ref{fig:evolution} and \ref{fig:teff_logg}.

If contact is made after core and shell He burning have finished (dashed grey and dashed-dotted black curves in Fig. \ref{fig:evolution}) and the sdB has become a C/O WD with a $0.41 M_\odot$ core and $0.06 M_\odot$ He envelope, the component that used to be the sdB will overflow its Roche lobe at an orbital period of less than 2 minutes, leading to a prompt merger and formation of an R\,CrB type star and subsequent evolution into a massive single WD.
Figure \ref{fig:evolution} shows that the post-CE orbital period lower limit for this outcome is 106\,min.

On the other hand, if the post-CE orbital period is less than 106 minutes, contact is made during the He burning phase, a merger may be avoided and the sdB will donate its remaining helium in an AM\,CVn type system \citep{bro15}.

\subsection{Current age}

\begin{figure}
\begin{center}
\includegraphics[width=0.49\textwidth]{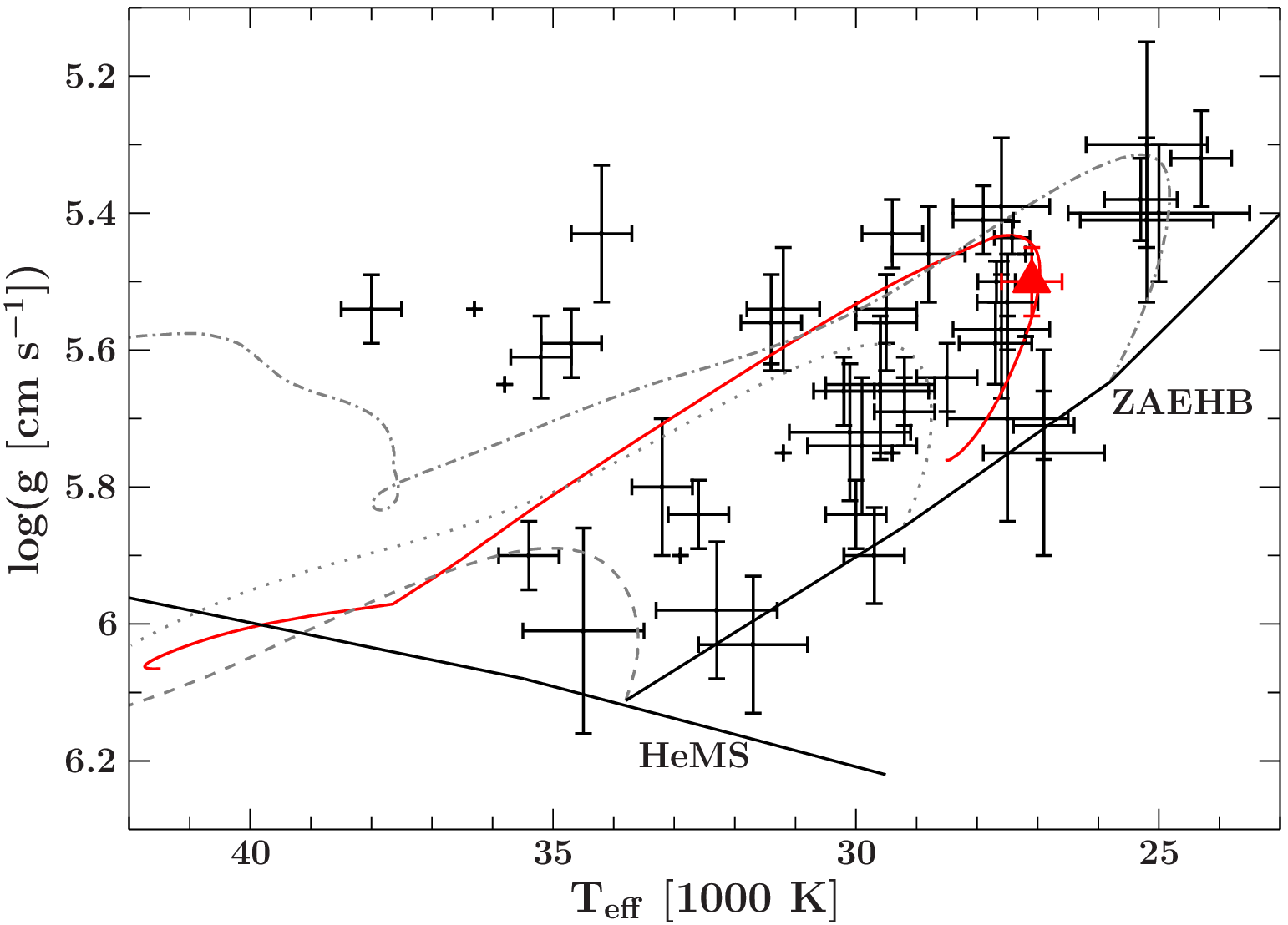}
\end{center}
\caption{\teff~-- \logg\,diagram of the compact binary sdB stars with confirmed white dwarf companions \citep{kup15a}. The red triangle marks \ptf. The helium main sequence (HeMS), the zero-age EHB (ZAEHB) and the terminal-age EHB (TAEHB) are superimposed with EHB evolutionary tracks by \citet{han02} (dashed lines: $m_{\rm env}=0.000$\,M$_\odot$, dotted lines: $m_{\rm env}=0.001$\,M$_\odot$, dashed-dotted lines: $m_{\rm env}=0.005$\,M$_\odot$ using $0.45$\,M$_\odot$ models). The red solid line shows the EHB evolutionary track calculated with \texttt{MESA} using a $0.47$\,M$_\odot$ model with $m_{\rm env}=5\times10^{-4}$\,M$_\odot$.}
\label{fig:teff_logg}
\end{figure}

Figure\,\ref{fig:teff_logg} shows the position of \ptf~ in the \teff~-- \logg~diagram overplotted with the confirmed sdB+white dwarf systems in compact orbits as well as theoretical evolutionary tracks. The sdBs with WD companions populate the full extreme horizontal branch (EHB) band homogeneously with a small fraction of sdBs having evolved off the EHB. The values of \teff~and \logg~for \ptf\, are consistent with an sdB on the EHB in the core helium burning stage.   

In a comparison of the spectroscopically derived $T_{\rm eff}$ and \logg\,with our \texttt{MESA} models, we find that an sdB model with a $5\times10^{-4}\,M_\odot$ hydrogen envelope matches the atmospheric parameter at a post-CE age of 94\,Myr (Fig.\,\ref{fig:teff_logg}), corresponding to a post-CE orbital period of 109\,min which is close to the limit where the sdB still burns helium when filling its Roche lobe.





\section{Conclusion and Summary}
Motivated by the possible existence of detached hot (ultra-)compact binaries such as CD$-$30$^\circ$11223, \ptf\,was found in a crossmatch between an SDSS color selected sample and the PTF database. 

The P48 lightcurve of \ptf, with a baseline of more than 5 years, revealed ellipsoidal deformation of the sdB. An orbital period of \porb=87.49668(1)\,min was found which makes \ptf\,the second-most compact sdB system known so far. Although the system is not eclipsing we have been able to derive a mass for the sdB of $M_{\rm sdB}=0.45^{+0.09}_{-0.07}$\,\msol\,and WD companion mass of $M_{\rm WD}=0.46^{+0.12}_{-0.09}$\,\msol\,{\bf by} assuming a tidally locked sdB. The distance was found to be $d=1.2^{+0.2}_{-0.2}$kpc. 

Although the solution allows for a wide range of companion masses we can exclude a massive white dwarf in \ptf:  $M_{\rm WD}<0.58$\,\msol. The upper limit on the WD mass is only possible if the sdB is also on the upper limit of its mass range with $M_{\rm sdB}\approx0.54$\,\msol\, which is not excluded, though only possible if the system evolved from a more massive binary system with main sequence components $>2$\,\msol. If the sdB has a canonical mass of $0.47$\,M$_\odot$ the companion is a low-mass white dwarf with a mass of $0.45<M_{\rm WD}<0.54$\,\msol.

\citet{kup15a} found that a significant fraction of the sdB binaries host WDs of masses below 0.6\,\msol\, but all have longer periods of at least a few hours. Therefore, \ptf\, is the first sdB with a confirmed low mass white dwarf companion in a tight orbit. 

We calculated the evolutionary history of \ptf\, using \texttt{MESA}, assuming a canonical sdB mass ($M_{\rm sdB}=0.47$\,M$_\odot$), and a companion mass of $M_{\rm WD}=0.49$\,\msol. We found that the sdB will burn helium for 152\,Myr and such a system will start accretion while the sdB is still burning helium if the orbital period after the system left the common envelope was smaller than 106\,min. 

If the system reaches contact after the helium burning has finished, the most likely outcome is a double white dwarf merger and subsequent evolution into an R\,CrB star with a mass of $0.8$ - $0.9 M_\odot$, which is the most common mass range for R\,CrB stars \citep{sai08,cla12}. The final evolutionary stage will be a single massive WD.

However, if the sdB still burns helium when the system reaches contact the sdB starts to accrete helium-rich material onto the WD companion and the most likely outcome is a helium accreting AM\,CVn type system. Therefore, compact sdB binaries with WD companions and post-CE orbital periods $\lesssim$100\,min might contribute to the population of AM\,CVn binaries with helium star donors. 

Whether \ptf\, is an R\,CrB progenitor or whether a merger is prevented and the system forms an AM\,CVn type system remains elusive and requires more detailed evolutionary calculations as well as more accurate mass measurements which will be available through \texttt{Gaia} photometry and parallaxes.


\section*{Acknowledgments}
This work was supported by the GROWTH project funded by the National Science Foundation under Grant No 1545949. JvR acknowledges support by the Netherlands Research School of Astronomy (NOVA) and the foundation for Fundamental Research on Matter (FOM). TRM acknowledge the support from the Science and Technology Facilities Council (STFC),  ST/L00733. This research was partially funded by the Gordon and Betty Moore Foundation through Grant GBMF5076 to Lars Bildsten. This work was supported by the National Science Foundation under grants PHY 11-25915, AST 11-09174, and AST 12-05574. This work was supported in part by the National Science Foundation under Grant No. PHYS-1066293 and the hospitality of the Aspen Center for Physics where parts of this paper was written. We thank the referee for helpful and timely comments.

Observations were obtained with the Samuel Oschin Telescope at the Palomar Observatory as part of the PTF project, a scientific collaboration between the California Institute of Technology, Columbia University, Las Cumbres Observatory, the Lawrence Berkeley National Laboratory, the National Energy Research Scientific Computing Center, the University of Oxford, and the Weizmann Institute of Science. Some of the data presented herein were obtained at the W.M. Keck Observatory, which is operated as a scientific partnership among the California Institute of Technology, the University of California and the National Aeronautics and Space Administration. The Observatory was made possible by the generous financial support of the W.M. Keck Foundation. The authors wish to recognize and acknowledge the very significant cultural role and reverence that the summit of Mauna Kea has always had within the indigenous Hawaiian community. We are
most fortunate to have the opportunity to conduct observations from this mountain. Some results presented in this paper are
based on observations made with the WHT operated on the island of La Palma by the Isaac Newton Group in the Spanish Observatorio del Roque de los Muchachos of the Institutio de Astrofisica de Canarias.

\software{\texttt{PyRAF} \citep{bel16}, \texttt{MAKEE} (http://www.astro.caltech.edu/∼tb/ipac staff/tab/makee/), \texttt{Gatspy} \citep{van15}, \texttt{LCURVE} \citep{cop10}, \texttt{emcee} \citep{for13}, \texttt{MESA} \citep{pax11,pax13,pax15}, \texttt{FITSB2} \citep{nap04a}}

\bibliographystyle{aasjournal}
\bibliography{refs}

\begin{thebibliography}{}
\expandafter\ifx\csname natexlab\endcsname\relax\def\natexlab#1{#1}\fi

\bibitem[{{Bellm} \& {Sesar}(2016)}]{bel16}
{Bellm}, E.~C., \& {Sesar}, B. 2016, {pyraf-dbsp: Reduction pipeline for the
  Palomar Double Beam Spectrograph}, Astrophysics Source Code Library, , ,
  ascl:1602.002

\bibitem[{{Bildsten} {et~al.}(2007){Bildsten}, {Shen}, {Weinberg}, \&
  {Nelemans}}]{bil07}
{Bildsten}, L., {Shen}, K.~J., {Weinberg}, N.~N., \& {Nelemans}, G. 2007, ApJL,
  662, L95

\bibitem[{{Bloemen} {et~al.}(2011){Bloemen}, {Marsh}, {{\O}stensen},
  {Charpinet}, {Fontaine}, {Degroote}, {Heber}, {Kawaler}, {Aerts}, {Green},
  {Telting}, {Brassard}, {G{\"a}nsicke}, {Handler}, {Kurtz}, {Silvotti}, {Van
  Grootel}, {Lindberg}, {Pursimo}, {Wilson}, {Gilliland}, {Kjeldsen},
  {Christensen-Dalsgaard}, {Borucki}, {Koch}, {Jenkins}, \& {Klaus}}]{blo11}
{Bloemen}, S., {Marsh}, T.~R., {{\O}stensen}, R.~H., {et~al.} 2011, \mnras,
  410, 1787

\bibitem[{{Brooks} {et~al.}(2015){Brooks}, {Bildsten}, {Marchant}, \&
  {Paxton}}]{bro15}
{Brooks}, J., {Bildsten}, L., {Marchant}, P., \& {Paxton}, B. 2015, \apj, 807,
  74

\bibitem[{{Carter} {et~al.}(1993)}]{car93}
{Carter}, D., {et~al.} 1993

\bibitem[{{Claret}(2004)}]{cla04}
{Claret}, A. 2004, \aap, 428, 1001

\bibitem[{{Clayton}(2012)}]{cla12}
{Clayton}, G.~C. 2012, Journal of the American Association of Variable Star
  Observers (JAAVSO), 40, 539

\bibitem[{{Constantino} {et~al.}(2015){Constantino}, {Campbell},
  {Christensen-Dalsgaard}, {Lattanzio}, \& {Stello}}]{con15}
{Constantino}, T., {Campbell}, S.~W., {Christensen-Dalsgaard}, J., {Lattanzio},
  J.~C., \& {Stello}, D. 2015, \mnras, 452, 123

\bibitem[{{Copperwheat} {et~al.}(2010){Copperwheat}, {Marsh}, {Dhillon},
  {Littlefair}, {Hickman}, {G{\"a}nsicke}, \& {Southworth}}]{cop10}
{Copperwheat}, C.~M., {Marsh}, T.~R., {Dhillon}, V.~S., {et~al.} 2010, \mnras,
  402, 1824

\bibitem[{{Dhillon} {et~al.}(2007){Dhillon}, {Marsh}, {Stevenson}, {Atkinson},
  {Kerry}, {Peacocke}, {Vick}, {Beard}, {Ives}, {Lunney}, {McLay}, {Tierney},
  {Kelly}, {Littlefair}, {Nicholson}, {Pashley}, {Harlaftis}, \&
  {O'Brien}}]{dhi07}
{Dhillon}, V.~S., {Marsh}, T.~R., {Stevenson}, M.~J., {et~al.} 2007, \mnras,
  378, 825

\bibitem[{{Dhillon} {et~al.}(2014){Dhillon}, {Marsh}, {Atkinson}, {Bezawada},
  {Bours}, {Copperwheat}, {Gamble}, {Hardy}, {Hickman}, {Irawati}, {Ives},
  {Kerry}, {Leckngam}, {Littlefair}, {McLay}, {O'Brien}, {Peacocke},
  {Poshyachinda}, {Richichi}, {Soonthornthum}, \& {Vick}}]{dhi14}
{Dhillon}, V.~S., {Marsh}, T.~R., {Atkinson}, D.~C., {et~al.} 2014, \mnras,
  444, 4009

\bibitem[{{Fink} {et~al.}(2010){Fink}, {R{\"o}pke}, {Hillebrandt},
  {Seitenzahl}, {Sim}, \& {Kromer}}]{fin10}
{Fink}, M., {R{\"o}pke}, F.~K., {Hillebrandt}, W., {et~al.} 2010, \aap, 514,
  A53

\bibitem[{{Foreman-Mackey} {et~al.}(2013){Foreman-Mackey}, {Hogg}, {Lang}, \&
  {Goodman}}]{for13}
{Foreman-Mackey}, D., {Hogg}, D.~W., {Lang}, D., \& {Goodman}, J. 2013, \pasp,
  125, 306

\bibitem[{{Geier} {et~al.}(2007){Geier}, {Nesslinger}, {Heber}, {Przybilla},
  {Napiwotzki}, \& {Kudritzki}}]{gei07}
{Geier}, S., {Nesslinger}, S., {Heber}, U., {et~al.} 2007, \aap, 464, 299

\bibitem[{{Geier} {et~al.}(2011){Geier}, {Hirsch}, {Tillich}, {Maxted},
  {Bentley}, {{\O}stensen}, {Heber}, {G{\"a}nsicke}, {Marsh}, {Napiwotzki},
  {Barlow}, \& {O'Toole}}]{gei11a}
{Geier}, S., {Hirsch}, H., {Tillich}, A., {et~al.} 2011, \aap, 530, A28

\bibitem[{{Geier} {et~al.}(2013){Geier}, {Marsh}, {Wang}, {Dunlap}, {Barlow},
  {Schaffenroth}, {Chen}, {Irrgang}, {Maxted}, {Ziegerer}, {Kupfer},
  {Miszalski}, {Heber}, {Han}, {Shporer}, {Telting}, {G{\"a}nsicke},
  {{\O}stensen}, {O'Toole}, \& {Napiwotzki}}]{gei13}
{Geier}, S., {Marsh}, T.~R., {Wang}, B., {et~al.} 2013, \aap, 554, A54

\bibitem[{{Han} {et~al.}(2003){Han}, {Podsiadlowski}, {Maxted}, \&
  {Marsh}}]{han03}
{Han}, Z., {Podsiadlowski}, P., {Maxted}, P.~F.~L., \& {Marsh}, T.~R. 2003,
  \mnras, 341, 669

\bibitem[{{Han} {et~al.}(2002){Han}, {Podsiadlowski}, {Maxted}, {Marsh}, \&
  {Ivanova}}]{han02}
{Han}, Z., {Podsiadlowski}, P., {Maxted}, P.~F.~L., {Marsh}, T.~R., \&
  {Ivanova}, N. 2002, \mnras, 336, 449

\bibitem[{{Heber}(1986)}]{heb86}
{Heber}, U. 1986, \aap, 155, 33

\bibitem[{{Heber}(2009)}]{heb09}
---. 2009, \araa, 47, 211

\bibitem[{Heber(2016)}]{heb16}
Heber, U. 2016, Publications of the Astronomical Society of the Pacific, 128,
  082001

\bibitem[{{Heber} {et~al.}(2000){Heber}, {Reid}, \& {Werner}}]{heb00}
{Heber}, U., {Reid}, I.~N., \& {Werner}, K. 2000, \aap, 363, 198

\bibitem[{{Henden} {et~al.}(2016){Henden}, {Templeton}, {Terrell}, {Smith},
  {Levine}, \& {Welch}}]{hen16}
{Henden}, A.~A., {Templeton}, M., {Terrell}, D., {et~al.} 2016, VizieR Online
  Data Catalog, 2336

\bibitem[{{Iben} \& {Tutukov}(1991)}]{ibe91}
{Iben}, Jr., I., \& {Tutukov}, A.~V. 1991, ApJ, 370, 615

\bibitem[{{Koen} {et~al.}(1998){Koen}, {Orosz}, \& {Wade}}]{koe98}
{Koen}, C., {Orosz}, J.~A., \& {Wade}, R.~A. 1998, \mnras, 300, 695

\bibitem[{{Kupfer} {et~al.}(2015){Kupfer}, {Geier}, {Heber}, {{\O}stensen},
  {Barlow}, {Maxted}, {Heuser}, {Schaffenroth}, \& {G{\"a}nsicke}}]{kup15a}
{Kupfer}, T., {Geier}, S., {Heber}, U., {et~al.} 2015, A\&A, 576, A44

\bibitem[{{Laher} {et~al.}(2014){Laher}, {Surace}, {Grillmair}, {Ofek},
  {Levitan}, {Sesar}, {van Eyken}, {Law}, {Helou}, {Hamam}, {Masci},
  {Mattingly}, {Jackson}, {Hacopeans}, {Mi}, {Groom}, {Teplitz}, {Desai},
  {Hale}, {Smith}, {Walters}, {Quimby}, {Kasliwal}, {Horesh}, {Bellm},
  {Barlow}, {Waszczak}, {Prince}, \& {Kulkarni}}]{lah14}
{Laher}, R.~R., {Surace}, J., {Grillmair}, C.~J., {et~al.} 2014, \pasp, 126,
  674

\bibitem[{{Law} {et~al.}(2009){Law}, {Kulkarni}, {Dekany}, {Ofek}, {Quimby},
  {Nugent}, {Surace}, {Grillmair}, {Bloom}, {Kasliwal}, {Bildsten}, {Brown},
  {et~al.}}]{law09}
{Law}, N.~M., {Kulkarni}, S.~R., {Dekany}, R.~G., {et~al.} 2009, PASP, 121,
  1395

\bibitem[{{Livne}(1990)}]{liv90}
{Livne}, E. 1990, \apjl, 354, L53

\bibitem[{{Livne} \& {Arnett}(1995)}]{liv95}
{Livne}, E., \& {Arnett}, D. 1995, \apj, 452, 62

\bibitem[{{Maxted} {et~al.}(2001){Maxted}, {Heber}, {Marsh}, \&
  {North}}]{max01}
{Maxted}, P.~f.~L., {Heber}, U., {Marsh}, T.~R., \& {North}, R.~C. 2001,
  \mnras, 326, 1391

\bibitem[{{Michaud} {et~al.}(2007){Michaud}, {Richer}, \& {Richard}}]{mic07}
{Michaud}, G., {Richer}, J., \& {Richard}, O. 2007, \apj, 670, 1178

\bibitem[{{Napiwotzki} {et~al.}(2004){Napiwotzki}, {Karl}, {Lisker}, {Heber},
  {Christlieb}, {Reimers}, {Nelemans}, \& {Homeier}}]{nap04a}
{Napiwotzki}, R., {Karl}, C.~A., {Lisker}, T., {et~al.} 2004, \apss, 291, 321

\bibitem[{{Nelemans}(2010)}]{nel10a}
{Nelemans}, G. 2010, \apss, 329, 25

\bibitem[{{Ofek} {et~al.}(2012){Ofek}, {Laher}, {Law}, {Surace}, {Levitan},
  {Sesar}, {Horesh}, {Poznanski}, {van Eyken}, {Kulkarni}, {Nugent},
  {Zolkower}, {Walters}, {Sullivan}, {Ag{\"u}eros}, {Bildsten}, {Bloom},
  {Cenko}, {Gal-Yam}, {Grillmair}, {Helou}, {Kasliwal}, \& {Quimby}}]{ofe12}
{Ofek}, E.~O., {Laher}, R., {Law}, N., {et~al.} 2012, \pasp, 124, 62

\bibitem[{{Oke} \& {Gunn}(1982)}]{oke82}
{Oke}, J.~B., \& {Gunn}, J.~E. 1982, PASP, 94, 586

\bibitem[{{Paxton} {et~al.}(2011){Paxton}, {Bildsten}, {Dotter}, {Herwig},
  {Lesaffre}, \& {Timmes}}]{pax11}
{Paxton}, B., {Bildsten}, L., {Dotter}, A., {et~al.} 2011, \apjs, 192, 3

\bibitem[{{Paxton} {et~al.}(2013){Paxton}, {Cantiello}, {Arras}, {Bildsten},
  {Brown}, {Dotter}, {Mankovich}, {Montgomery}, {Stello}, {Timmes}, \&
  {Townsend}}]{pax13}
{Paxton}, B., {Cantiello}, M., {Arras}, P., {et~al.} 2013, \apjs, 208, 4

\bibitem[{{Paxton} {et~al.}(2015){Paxton}, {Marchant}, {Schwab}, {Bauer},
  {Bildsten}, {Cantiello}, {Dessart}, {Farmer}, {Hu}, {Langer}, {Townsend},
  {Townsley}, \& {Timmes}}]{pax15}
{Paxton}, B., {Marchant}, P., {Schwab}, J., {et~al.} 2015, \apjs, 220, 15

\bibitem[{{Piersanti} {et~al.}(2014){Piersanti}, {Tornamb{\'e}}, \&
  {Yungelson}}]{pie14}
{Piersanti}, L., {Tornamb{\'e}}, A., \& {Yungelson}, L.~R. 2014, \mnras, 445,
  3239

\bibitem[{{Ramspeck} {et~al.}(2001){Ramspeck}, {Heber}, \& {Edelmann}}]{ram01}
{Ramspeck}, M., {Heber}, U., \& {Edelmann}, H. 2001, \aap, 379, 235

\bibitem[{{Rau} {et~al.}(2009){Rau}, {Kulkarni}, {Law}, {Bloom}, {Ciardi},
  {Djorgovski}, {Fox}, {Gal-Yam}, {Grillmair}, {Kasliwal}, {Nugent}, {Ofek},
  {Quimby}, {Reach}, {Shara}, {Bildsten}, {Cenko}, {Drake}, {Filippenko},
  {Helfand}, {Helou}, {Howell}, {Poznanski}, \& {Sullivan}}]{rau09}
{Rau}, A., {Kulkarni}, S.~R., {Law}, N.~M., {et~al.} 2009, \pasp, 121, 1334

\bibitem[{{Saio}(2008)}]{sai08}
{Saio}, H. 2008, in Astronomical Society of the Pacific Conference Series, Vol.
  391, Hydrogen-Deficient Stars, ed. A.~{Werner} \& T.~{Rauch}, 69

\bibitem[{{Savonije} {et~al.}(1986){Savonije}, {de Kool}, \& {van den
  Heuvel}}]{sav86}
{Savonije}, G.~J., {de Kool}, M., \& {van den Heuvel}, E.~P.~J. 1986, A\&A,
  155, 51

\bibitem[{{Schindler} {et~al.}(2015){Schindler}, {Green}, \& {Arnett}}]{sch15}
{Schindler}, J.-T., {Green}, E.~M., \& {Arnett}, W.~D. 2015, \apj, 806, 178

\bibitem[{{Shakura} \& {Postnov}(1987)}]{sha87}
{Shakura}, N.~I., \& {Postnov}, K.~A. 1987, \aap, 183, L21

\bibitem[{{Shen} \& {Bildsten}(2014)}]{she14}
{Shen}, K.~J., \& {Bildsten}, L. 2014, \apj, 785, 61

\bibitem[{{Tutukov} \& {Fedorova}(1989)}]{tut89}
{Tutukov}, A.~V., \& {Fedorova}, A.~V. 1989, Soviet Astronomy, 33, 606

\bibitem[{{Tutukov} \& {Yungelson}(1990)}]{tut90}
{Tutukov}, A.~V., \& {Yungelson}, L.~R. 1990, \sovast, 34, 57

\bibitem[{{VanderPlas} \& {Ivezi\'{c}}(2015)}]{van15}
{VanderPlas}, J.~T., \& {Ivezi\'{c}}, v. 2015, \apj, 812, 18

\bibitem[{{Vennes} {et~al.}(2012){Vennes}, {Kawka}, {O'Toole}, {N{\'e}meth}, \&
  {Burton}}]{ven12}
{Vennes}, S., {Kawka}, A., {O'Toole}, S.~J., {N{\'e}meth}, P., \& {Burton}, D.
  2012, \apjl, 759, L25

\bibitem[{{Verbunt} \& {Rappaport}(1988)}]{ver88}
{Verbunt}, F., \& {Rappaport}, S. 1988, \apj, 332, 193

\bibitem[{{von Zeipel}(1924)}]{zei24}
{von Zeipel}, H. 1924, \mnras, 84, 665

\bibitem[{{Wang} {et~al.}(2009){Wang}, {Meng}, {Chen}, \& {Han}}]{wan09}
{Wang}, B., {Meng}, X., {Chen}, X., \& {Han}, Z. 2009, \mnras, 395, 847

\bibitem[{{Woosley} \& {Kasen}(2011)}]{woo11}
{Woosley}, S.~E., \& {Kasen}, D. 2011, \apj, 734, 38

\bibitem[{{Yungelson}(2008)}]{yun08}
{Yungelson}, L.~R. 2008, Astronomy Letters, 34, 620

\end{thebibliography}

\end{document}